\documentclass[sigconf, 10pt, nonacm, screen]{acmart}

\usepackage{graphicx} 

\usepackage[utf8]{inputenc} 
\usepackage{url}            
\usepackage{booktabs}       
\usepackage{amsfonts,amsmath}       
\usepackage{nicefrac}       
\usepackage{microtype}      
\usepackage{color}
\usepackage{graphicx}
\usepackage{xcolor}         
\usepackage{colortbl}       
\usepackage{fixltx2e}
\usepackage{subcaption}
\usepackage{cleveref}
\usepackage{xspace}

\usepackage{hyperref}
\usepackage{paralist}
\usepackage{subcaption} 
\usepackage{tabularx}

\newif\ifshowcomments
\showcommentsfalse  

\makeatletter
\@ifundefined{xspace@exceptions}{}{
  \g@addto@macro\xspace@exceptions{'`}
}
\makeatother
\DeclareRobustCommand{\toolname}{RFSeek\xspace}

\newcommand{\circled}[1]{\textcircled{\raisebox{-.9pt} {#1}}}

\title{\toolname and Ye Shall Find}
\subtitle{A tool for summary visualization and analysis of RFCs}

\author{Noga H. Rotman}
\affiliation{%
    \institution{Technion - Israel Institute of Technology}
     \country{}
}
\email{noga.rotman@mail.huji.ac.il}

\author{Tiago Ferreira}
\affiliation{%
    \institution{University College London}
     \country{}
}
\email{t.ferreira@ucl.ac.uk}

\author{Hila Peleg}
\affiliation{%
    \institution{Technion - Israel Institute of Technology}
     \country{}
}
\email{hilap.cs@gmail.com}

\author{Mark Silberstein}
\affiliation{%
    \institution{Technion - Israel Institute of Technology}
     \country{}
}
\email{mark@ee.technion.ac.il}

\author{Alexandra Silva}
\affiliation{%
    \institution{Cornell University}
     \country{}
}
\email{alexandra.silva@cornell.edu}

\makeatletter
\def\@acmplainindent{0pt}
\def\@acmdefinitionindent{0pt}
\def\@proofindent{\noindent}
\makeatother

\newcommand{\inlinehead}[1]{\par\noindent\textit{#1.}\ \ignorespaces}

\usepackage[skip=2pt]{caption} 
\begin{document}

\begin{abstract}
Requests for Comments (RFCs) are extensive specification documents for network protocols, but their prose-based format and their considerable length often impede precise operational understanding. 
We present \toolname, an interactive tool that automatically extracts visual summaries of protocol logic from RFCs. \toolname leverages large language models (LLMs) to generate provenance-linked, explorable diagrams, surfacing both official state machines and additional logic found only in the RFC text. Compared to existing RFC visualizations, \toolname's visual summaries are more transparent and easier to audit against their textual source.
We showcase the tool's potential through a series of use cases, including guided knowledge extraction and semantic diffing, applied to protocols such as TCP, QUIC, PPTP, and DCCP. 

In practice, \toolname not only reconstructs the RFC diagrams included in some specifications, but, more interestingly, also uncovers important logic such as nodes or edges described in the text but missing from those diagrams. 
\toolname further derives new visualization diagrams for complex RFCs, with QUIC as a representative case. 
Our approach, which we term \emph{Summary Visualization}, highlights a promising direction: combining LLMs with formal, user-customized visualizations to enhance protocol comprehension and support robust implementations.
\end{abstract}

\maketitle

\section{Introduction}


Requests for Comments (RFCs) are the documents providing the authoritative standards for fundamental Internet protocols (e.g. TCP, HTTP, DNS, etc.), and serve as the definitive source for understanding and implementation guidance. 
New protocols begin their life cycle as \textit{living documents}, dozens of pages long, written by human authors in English. As such, they are prone to omissions and inconsistencies, that can appear many pages apart. 
In addition, their descriptions are intentionally permissive, in order to allow for general interfaces and different ``flavoring'' when put to practice, further contributing to 
textual ambiguity.
Authors of RFCs often include figures to assist implementers, but these depictions of Finite State Machines (FSMs) are typically abstract, and often incomplete.

To bridge the gap between RFC authors and protocol developers,
we propose \toolname, an interactive and malleable 
visualization of protocol states and governing events.
We term this approach \emph{Summary Visualization}, in which a summary of protocol behavior is automatically produced from the RFC and transformed into a provenance-aware visual model. Underlying the tool is an automated framework leveraging Large Language Models (LLMs), which the user is agnostic to, while ensuring every diagram element is grounded in specific RFC content.

Let us take the case of an early developer working on a new version of the TCP RFC (9293).

\begin{figure}[!h]    
\includegraphics[width=0.56\columnwidth]{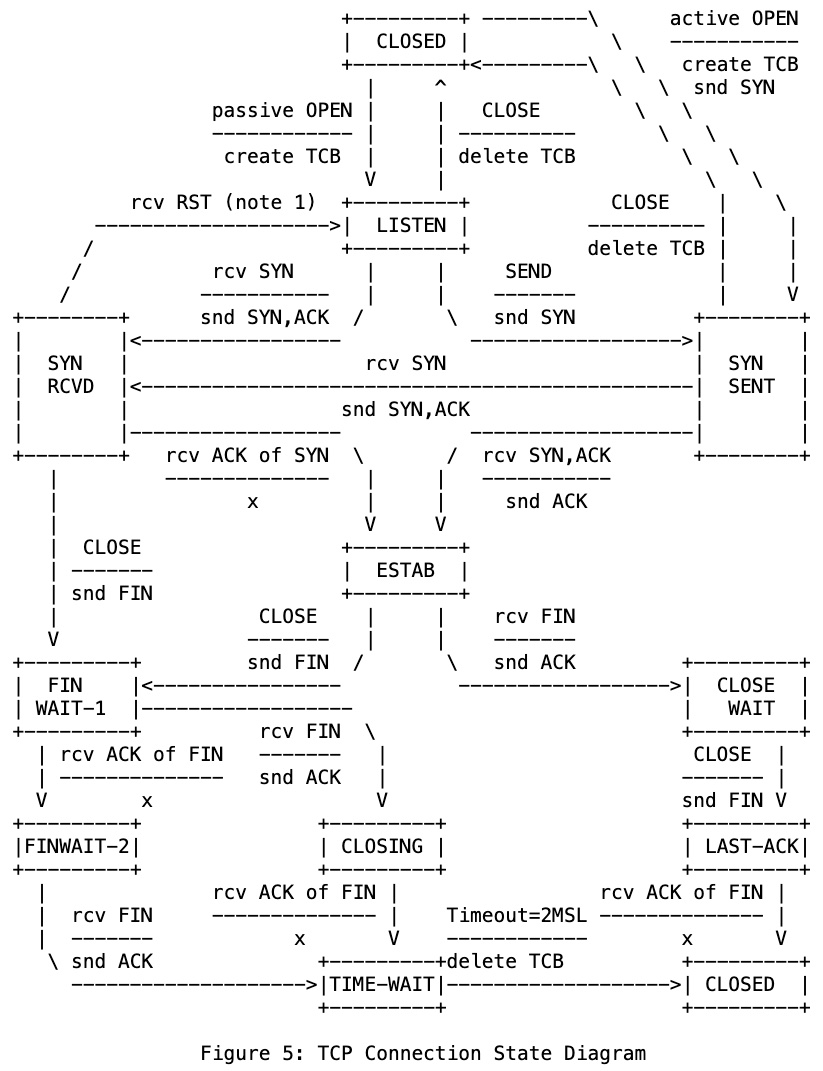}
    \caption{ASCII diagram from RFC 9293}\label{fig:tcp-rfc-asii}
\end{figure}

\begin{figure*}[!t]
    \begin{subfigure}{0.4\textwidth}
\hspace*{-.85cm}\includegraphics[width=1.01\columnwidth]{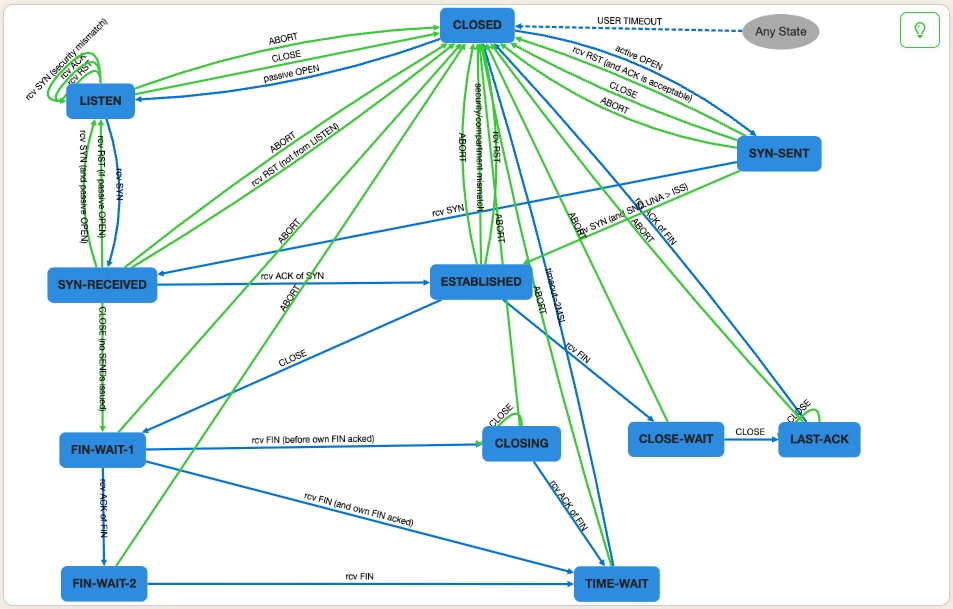}
        \caption{}
        \label{fig:tool-full}
    \end{subfigure}
    \begin{subfigure}{0.5\textwidth}        \hspace*{-.49cm}\includegraphics[width=1.15\columnwidth]{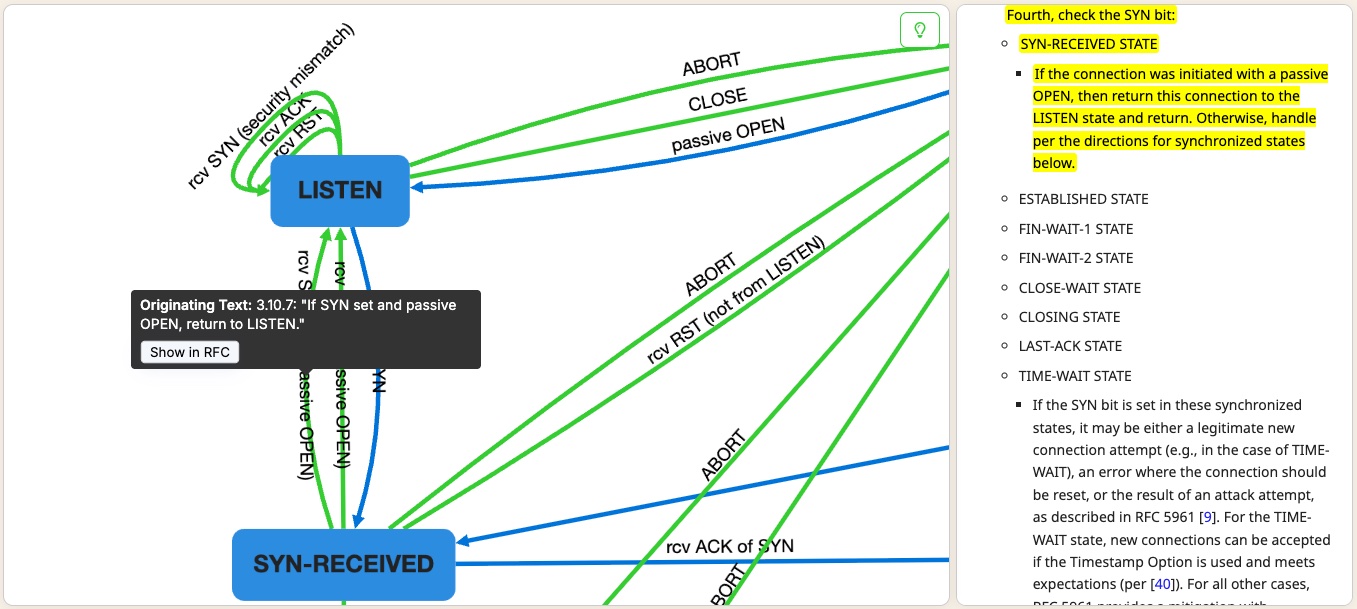}
        \caption{}
        \label{fig:new-edge}
    \end{subfigure}

\caption{\toolname analysis of TCP (RFC 9293): On the left, the summary of RFC 9293 created by \toolname. Grey oval nodes are \emph{any} nodes. On the right, we zoom in on a new edge found by using \toolname, from \texttt{SYN-RECEIVED} to \texttt{LISTEN}. The edge is green, indicating it not being a part of the description of the FSM in the RFC text, but stemming from another section of the text. Hovering over the edge shows the text that caused that edge to be created. }
\end{figure*}
They might start their implementation by looking at the ASCII-art diagram 
in Section 3.3.2 of the RFC, 
shown in \Cref{fig:tcp-rfc-asii}.
This diagram is inherently incomplete because it is an abstraction,
and the RFC is explicit about this, clearly stating that there are missing edges.



Seeking a more complete picture, our developer turns to \toolname and loads the TCP RFC into the tool. \toolname present them with the summary visualization shown in \Cref{fig:tool-full}.
While some of the edges are easily recognizable---%
even from an undergraduate networking class---%
our developer inspects the less familiar transitions, which are precisely the ones that motivated them to use \toolname in the first place.  
%
They hone in on an edge from the \texttt{SYN-RECEIVED} state
to the \texttt{LISTEN} state, created by \texttt{rcv SYN}.
Note that this edge is absent from both the ASCII diagram and its accompanying notes in the RFC, but it describes the stated behavior in case a \texttt{SYN} is received after an initial one was already received in a \texttt{passive OPEN} connection, leading to the \texttt{SYN-RECEIVED} state. 

The developer then hovers on the transition to see additional details,
 clicking on the ``show in RFC'' button to inspect the text 
from the RFC that created the transition in \toolname (\Cref{fig:new-edge}).
After validating the transition (explained in §3.10.7.4 of the RFC)
and confirming that the clarifications below the ASCII diagram do not mention its omission,
they send a note to the RFC authors.
At the very least, they suggest adding a reference to the omission in Section 3.3.2.

In short,
\toolname aids in 
a deeper understanding of the protocol,
and may facilitate feedback from developers to RFC authors,
contributing to improved specifications and implementations of RFCs.



\paragraph{Previous approaches}
Numerous prior works have tackled extracting models from RFC text, but
\toolname differs on two significant fronts.
First, the core problem \toolname addresses is fundamentally different.
Most tools are intended for automated tasks like fuzzing~\cite{jero_leveraging_2019, zhang_blackbox_2023, meng_large_2024} or attack synthesis~\cite{pacheco_automated_2022}, where some inaccuracy is tolerable and might even be beneficial (e.g., more transitions can aid coverage in fuzzing).
In contrast, \toolname is designed to advance the accurate understanding of RFCs and ultimately support improved protocol specifications. 
As a result, our approach demands outputs that are as correct and faithful to the source as possible. 
Second, and not unrelated, existing tools typically produce FSMs
with
only the minimal information needed for automatic traversal, 
omitting many of the nuances present in protocol transitions.
\toolname, on the other hand, generates diagrams that are both comprehensive and readable, capturing richer context, transition rationale, and provenance information---%
features critical for in-depth analysis and auditability.



\toolname also stands out in its extraction technique.
Most importantly,
\toolname does not rely on the ASCII-art diagrams in the RFC~\cite{sharma_prosper_2023},
which, if present, are inherently incomplete. 
Such diagrams can also be difficult to parse automatically, introducing errors or omissions into extracted models.
Moreover, 
unlike several recent approaches, \toolname
does not require model training~\cite{jero_leveraging_2019, pacheco_automated_2022}, nor does it rely on the users to supply supplementary resources such as a technical lexicon~\cite{yen_semi-automated_2021}. This makes \toolname both easier to deploy and more broadly applicable across diverse protocol documents.
Our processing techniques set a new standard for auditability in protocol analysis: every extracted element is explicitly and transparently grounded in the RFC text.

In a nutshell, our {contributions} are:
\begin{compactenum}[1.]
    \item \textbf{A new, principled summary representation for RFC protocols} that extends and defines the state machine information traditionally conveyed via informal diagrams.
    \item \textbf{A modular extraction workflow} that transforms RFC documents into structured, explorable summaries while retaining both long-range cross-references and traceability to source text.
    \item \textbf{\toolname, an interactive environment for protocol summary exploration}, enabling users to directly trace each transition to its RFC source, and supporting protocols regardless of whether FSM diagrams are comprehensive, fragmented, or absent, as exemplified by QUIC, which includes only partial figures and  text-only state machines.

    \item \textbf{A preliminary evaluation} of our prompting strategy and extraction method, including a comparative analysis between \toolname-generated summaries and the ASCII diagrams from several RFCs---TCP, QUIC, DCCP and PPTP.
\end{compactenum}

\section{Approach}\label{sec:methodology}
In this section, we describe the components of our framework: the  summary representation, the pipeline of the tool, and the prompting strategy we use to maximize automation. 
\subsection{Proposed Summary Definition}\label{subsec:summary_def}
Some RFCs include one or more FSM diagrams. However, because the official RFC is published as a plain text document, all diagrams, including FSMs, are restricted to ASCII-art representations. As a result, these figures provide only partial information. 
For example, the official TCP FSM ( \Cref{fig:tcp-rfc-asii}) is accompanied by a disclaimer: ``many details are not included'', and additional transitions are omitted lest ``the diagram would become very difficult to read''.

In this work, our goal is to capture more of the rich tapestry of protocol behavior described in RFCs. Rather than relying on the varied and often informal FSM definitions found across RFCs and previous extraction efforts (see \Cref{sec:related}), we introduce a new summary representation that formalizes and unifies protocol behavior as presented in both diagrams and text.
First, by design, our summary does not omit transitions it learned grounded on evidence.  
However, when multiple states share the same event and handle it identically, we introduce a representative grouped node
to avoid overcrowding the presentation. 
Second, we include transitions that are \emph{recommended but are not mandatory}, and transitions that are \emph{inferred} from the text. In the case of inferred transitions, our summary always includes the reasoning for their creation.
To the best of our knowledge, this is the first time such edges have been introduced.
Finally, while other FSMs may include partial information regarding each edge, in our summary, each transition is described by: \begin{inparaenum}[(i)]
\item the triggering event, and any relevant conditions;
\item the action that should be taken, if any, in detail - including the construction / destruction of data structures, error codes, and any other pertinent information;
\item the originating text (see §\ref{subsec:pipeline}), and
\item in case of a grouped transition: which states are included.
\end{inparaenum}


\subsection{Pipeline}\label{subsec:pipeline}
\begin{figure}[t]
    \centering
    \includegraphics[width=.8\linewidth]{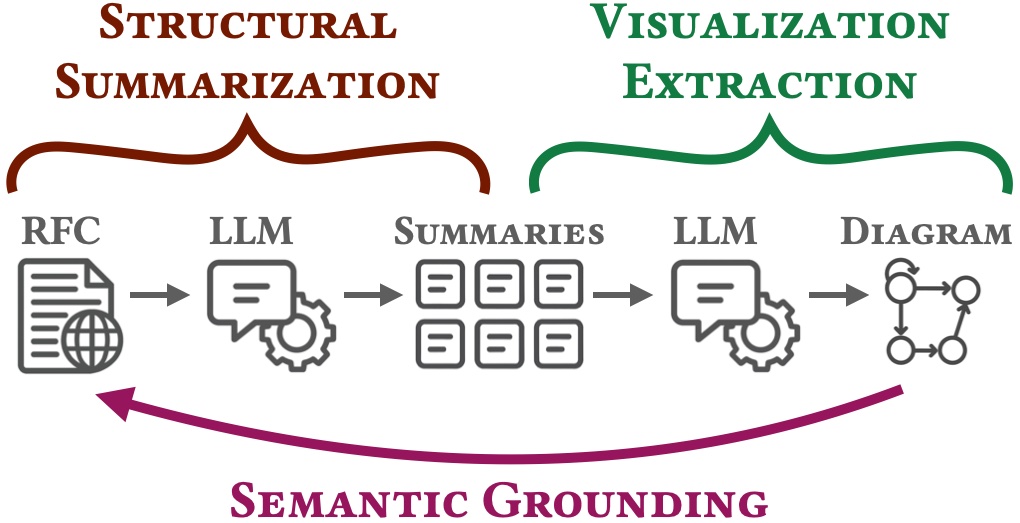}
    \caption{The pipeline of \toolname}
    \label{fig:pipeline}
\end{figure}

Effective use of LLMs requires balancing specificity and clarity in prompts, as overly detailed instructions can solicit unintelligible answers.
\Cref{fig:pipeline} depicts our pipeline, partitioned
into specific tasks that LLMs excel at, such as summarization and semantic grounding~\cite{rag}. All experiments used the OpenAI GPT-4.1 model via the OpenAI API.

RFC documents are typically too large to be processed by LLMs in a single pass due to input length restrictions. 
We therefore begin by 
dividing the RFC into \emph{structural} components such as sections, subsections or even smaller fragments, depending on their size.
Prior to partitioning, we apply standard preprocessing steps (such as whitespace normalization and ASCII table condensation) to improve input quality and consistency. 
For each component, we compute dense vector representations (embeddings) and use these to retrieve semantically relevant excerpts from elsewhere in the document to supplement the main content during summarization. This ensures the LLM receives not only a well-defined portion of the text, but also any pertinent supporting information, improving the accuracy and level of detail in the summaries.
The summaries we produce are designed to align with our proposed summary definition (see \Cref{subsec:summary_def}), ensuring that the extracted information faithfully represents the protocol semantics as captured by our approach.

Next, we prompt the LLM to perform the \emph{visualization extraction}. For each edge, we instruct the LLM to identify and cite the specific summary segment(s) that serve as the basis for that transition.
Using summary-based input enables us to include the full RFC context within a single prompt.
%

In the final step, \emph{semantic grounding}, we prompt the LLM to retrieve the corresponding RFC text passages that justify each edge.
The full visualization summary is then loaded into our user interface (detailed in~\Cref{sec:ui}) 
allowing users to explore the constituent nodes and edges.

\subsection{Prompting Strategy}\label{subsec:prompting}
We explored several strategies before settling on our current methodology. As a baseline, we first tested prompting the LLM with only those RFC sections that were deemed most relevant, to assess whether selective input could efficiently recover meaningful protocol transitions. While this sanity check performed reasonably well, it reproduced the transitions already depicted in the diagrams and did not yield any new or implicit protocol behaviors.

Next, we introduced a summarization stage prior to visualization extraction. 
However, we found that while some sections could be summarized in their entirety, it was rare for an RFC to contain only such manageable sections. In most cases, at least some sections required further partitioning into smaller fragments to fit input constraints and maintain semantic coherence.

We also compared general-purpose summaries to targeted summaries focused on FSM extraction---%
that is, prompts specifically designed to elicit the transitions and context captured by our summary definition. 
Using shorter, targeted summaries did not reduce precision, so we adopted them as our default input.

To further probe the LLM’s reliance on textual context, we evaluated the effect of omitting the original FSM diagram from the RFC input (particularly for RFC9293). 
When the diagram was absent, certain transitions were missing from the summaries, as they were not directly mentioned elsewhere in the document. While we did not systematically assess this across all protocols, this suggests that 
transitions described exclusively in diagrams may be overlooked by LLM-based extraction methods focused on text.

To ensure traceability, we required that every extracted transition be explained by its originating text. 
Notably, when we prompted the LLM to extract a ``precise and accurate FSM'' it completely omitted some edges it had previously identified, such as the one shown in \Cref{fig:new-edge}.

A related observation was that some of the transitions clearly mentioned in the section summaries were missing from the extracted visualization summary. To address this, we adjusted our prompt to explicitly instruct the LLM to extract all transitions mentioned in the summaries. Interestingly, this did not increase the total number of transitions identified; rather, the set of extracted transitions shifted. With the revised prompt, only transitions explicitly mentioned in the summaries were extracted, while transitions previously inferred implicitly by the LLM from the text were now omitted. This warrants further investigation.

\section{User Interface}\label{sec:ui}

\begin{figure}[t]
\begin{subfigure}{\columnwidth}
\centering
\includegraphics[width=0.98\columnwidth]{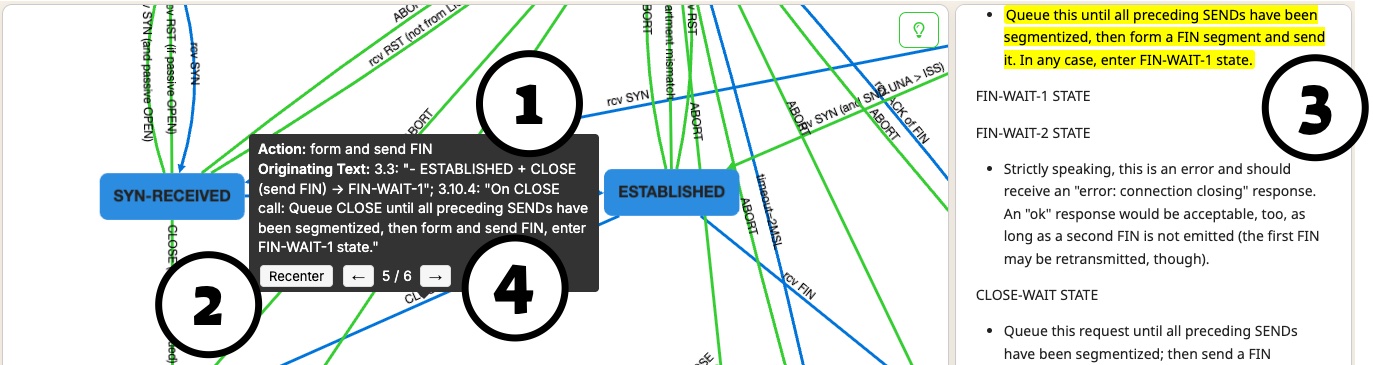}
\caption{Hovering on an edge in \toolname, showing 
$\circled{1}$ summary text that created the edge,
$\circled{2}$ a ``Recenter'' button to scroll back to the 
$\circled{3}$ highlighted RFC text, and an
$\circled{4}$ indicator this is RFC location 5 of 6 expressing the current edge. Arrows will move to the previous/next location in the text.}\label{fig:multiplesrcs}
\end{subfigure}
\begin{subfigure}{\columnwidth}
\centering
\includegraphics[width=0.88\columnwidth]{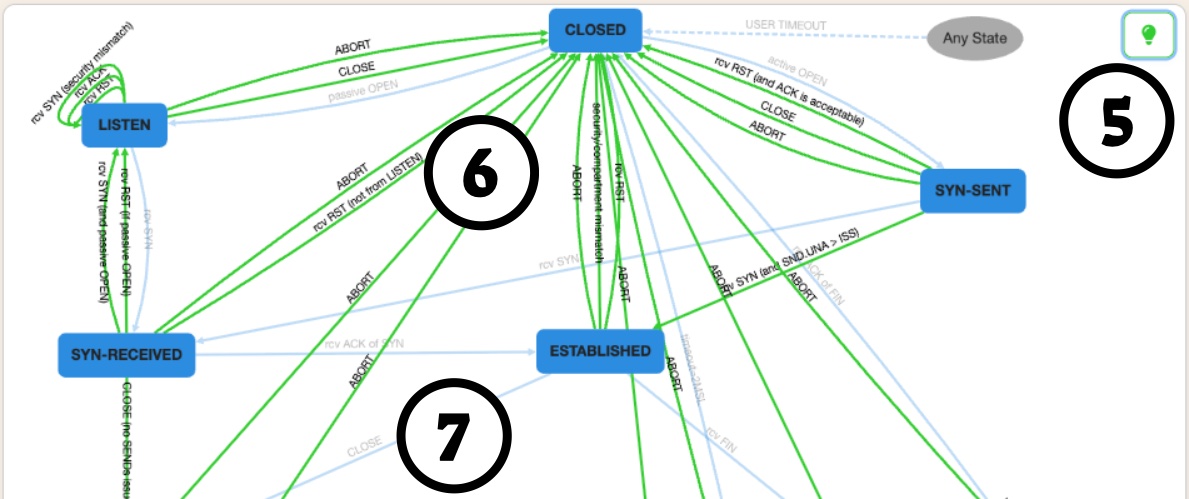}
\caption{Toggling the $\circled{5}$ light bulb button highlights $\circled{6}$ edges that are only described in the text and grays out $\circled{7}$ edges from the ASCII diagram.}\label{fig:lightbulb}
\end{subfigure}
\caption{Inspection features in \toolname}
\end{figure}

Once an RFC is processed, the resulting summary visualization is
loaded in the \toolname UI.

\inlinehead{Inspecting summary visualizations}
The user can zoom in and out on different parts of the diagram, and 
freely reposition nodes and edges to reorganize the visualization as needed.
Node and edge labels are editable for clarity or annotation.
To make the interface reusable, any user customizations can be saved and reloaded in future sessions.

\inlinehead{Navigating the RFC}
The user can hover over any edge to view the specific summary excerpt(s) 
($\circled{1}$ in \Cref{fig:multiplesrcs}).
Clicking the ``Show in RFC'' button (see \Cref{fig:new-edge}) automatically scrolls the RFC side panel ($\circled{3}$ in \Cref{fig:multiplesrcs}) to the first relevant passage and highlights all supporting locations.
The button then changes to ``Recenter'' ($\circled{2}$), letting the user browse freely and 
return to the source with a single click.

If an edge is justified by multiple text passages, the summary tooltip displays progress (e.g., ``5/6'' $\circled{4}$) and arrow buttons enable navigation between all supporting RFC snippets.

\inlinehead{Highlighting new edges}
Summary visualizations of large RFCs can contain many edges, 
some originating from the main FSM diagram section and others from elsewhere in the RFC text. 
\toolname colors edges from the FSM section blue (see \Cref{fig:tool-full}) and highlights edges from other RFC sections in green.

To help the user focus on new edges sourced from the broader text, 
the light bulb button ($\circled{5}$ in \Cref{fig:lightbulb})
toggles a view that grays out diagram-section edges ($\circled{7}$) and highlights those extracted from other sections ($\circled{6}$).

\section{Case Studies}\label{sec:eval}


To evaluate \toolname, we applied it to four protocols:
PPTP (RFC2637), DCCP (RFC4341), QUIC (RFC9000), and TCP (RFC9293).
For each, we measured how faithfully \toolname recovered FSM edges and nodes depicted in the published diagrams, and identified additional protocol logic surfaced from the text.
While we do not claim completeness, we focus on correctness:
our extracted summary diagrams are grounded in and traceable to the RFC source.
We also compare \toolname's results to those of 
PROSPER~\cite{sharma_prosper_2023} 
(see \Cref{tab:rfc_methods_two}).

\begin{table}[t]
\centering
\caption{Comparison of missing nodes and edges (lower is better) across RFCs for PROSPER~\cite{sharma_prosper_2023} and RFSeek.
}
\label{tab:rfc_methods_two}
\begin{tabularx}{\columnwidth}{l *{2}{>{\centering\arraybackslash}X} *{2}{>{\centering\arraybackslash}X}}
\toprule
RFC 
  & \multicolumn{2}{c}{PROSPER} 
  & \multicolumn{2}{c}{RFSeek} \\
  \cmidrule(lr){2-3} \cmidrule(lr){4-5}
  & Missing Nodes & Missing Edges
  & Missing Nodes & Missing Edges \\
\midrule
PPTP (RFC2637)  & 0 & 19 & 0 & 6 \\
DCCP (RFC4341)  & 1 & 7 & 0 & 1 \\
QUIC (RFC9000)$^1$  & - & - & 0 & 2 \\ 
TCP (RFC9293)$^2$   & - & - & 0 & 1 \\ 
\bottomrule
\multicolumn{5}{l}{{\footnotesize 
${}^1$\emph{The RFC only shows two diagrams, others are described,
making it a poor}}}\\ 
\multicolumn{5}{l}{{\footnotesize \emph{
candidate for PROSPER.}}}\\
\multicolumn{5}{l}{{\footnotesize
${}^2$\emph{PROSPER use RFC 793 for TCP, making a direct comparison unfair.}
}}
\end{tabularx}
\vspace{-.3cm}
\end{table}

\subsection{Case study: TCP}
TCP~\cite{rfc9293} is among the most prominent and widely deployed networking protocols. RFC9293 consolidates over forty years of evolution and multiple updates into a single, unified specification.
Given this, one might expect the FSM of TCP would be agreed upon. Yet, we were surprised \toolname identified a new transition absent from both the FSM diagram and its accompanying notes.
This edge, depicted in \Cref{fig:new-edge}, captures the transition from \texttt{SYN-RECEIVED} to \texttt{LISTEN}, which occurs upon receiving a \texttt{SYN} provided that the connection was initialized with \texttt{passive OPEN}.  Otherwise, the protocol specifies a different handling.
Interestingly, a review of the Linux kernel's TCP implementation~\cite{ferreira2021prognosis} reveals that this transition is, in fact, present there.

\subsection{Case study: PPTP}
PPTP (RFC2637) defines its state machine through six separate diagrams.
\toolname recovered all but six diagrammed transitions, while also surfacing new edges and a previously undocumented node not present in any diagram.

\begin{figure}[h!]
    \begin{subfigure}{\columnwidth}
    \centering
    \includegraphics[width=0.98\columnwidth]{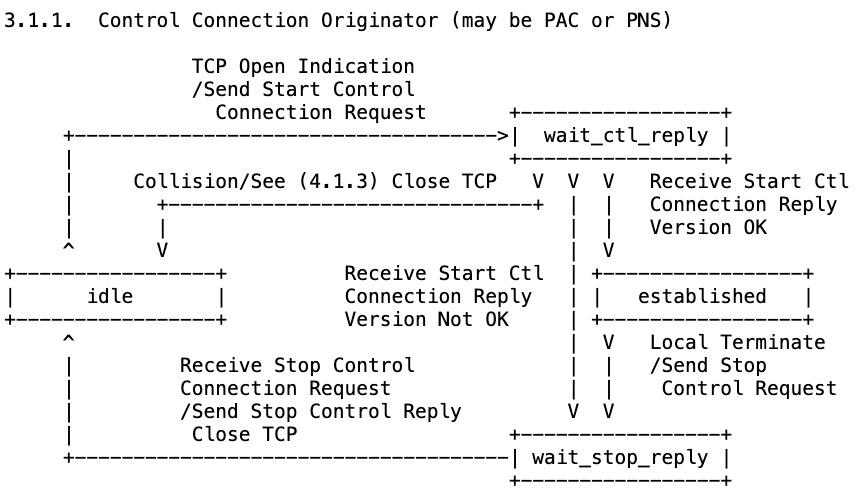}
    \caption{ASCII diagram from RFC2637 PPTP. Note the edge from \texttt{wait\_ctl\_reply} to \texttt{idle}. Section §3.1.3 of the RFC (4.1.3 does not exist) clearly states that this edge exists only for the "loser" of the initiation race, the "winner" should continue, and specifically does not return to \texttt{idle}.}
    \label{fig:pptp_rfc-ascii}
    \end{subfigure}

    \begin{subfigure}{\columnwidth}
        \centering
        \includegraphics[width=0.94\columnwidth]{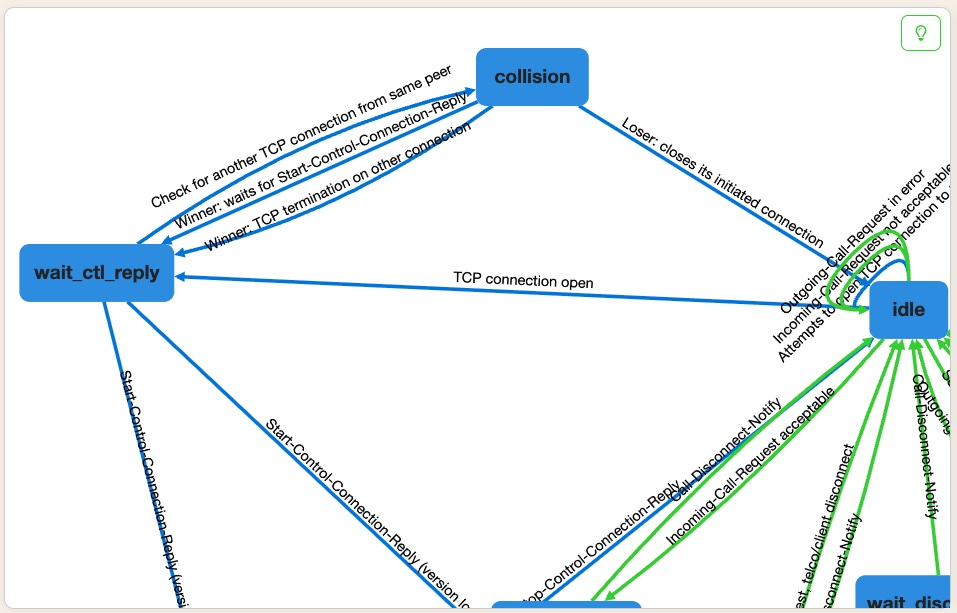}
        \caption{The new node constructed by \toolname, \texttt{collision}. Note the edge from \texttt{collision} to \texttt{idle} for the losing party, and two edges to \texttt{wait\_ctl\_reply} for the winner.}
        \label{fig:pptp_new_node}
    \end{subfigure}
\caption{PPTP (RFC2637) in the RFC and in \toolname}\label{fig:pptp}
\end{figure}

Consider the transition from \texttt{wait\_ctl\_reply} to \texttt{idle}, described in the RFC diagram (see \Cref{fig:pptp_rfc-ascii}). Section 3.1.3 of the RFC clearly states that in case of a collision, ``The loser will immediately close the TCP connection it initiated''. This is indicated by the edge described above. However, the text also specifies what should happen to the connection the winner initiated, which will not terminate, but instead return to \texttt{wait\_ctl\_reply} given specific conditions. This example demonstrates \toolname's ability to synthesize details from the RFC text that are not captured in the official diagrams, enabling a more complete and accurate summary.


\subsection{Case study: QUIC}
\begin{figure}[h!]
    \centering
    \includegraphics[width=0.94\columnwidth]{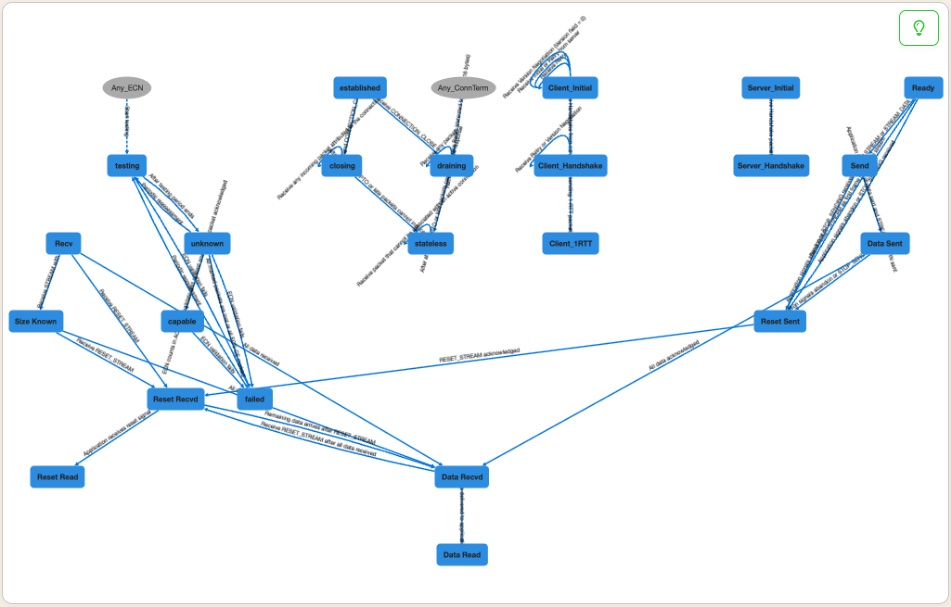}
    \caption{QUIC (RFC9000) in \toolname}\label{fig:quic}
\end{figure}

The relatively new QUIC protocol~\cite{quic} has already had a significant impact on both academic research and industry deployment. 
Although its RFC provides partial and inconsistent state machine figures for select procedures (e.g. Streams in §3.1-2), \toolname not only reconstructs and unifies these visualizations, showcasing their interactions, but also surfaces additional procedures not visualized in the official document (see \Cref{fig:quic}). 
This yields a more unified and complete view of protocol logic than the RFC diagrams alone.


\subsection{Case study: DCCP}
For DCCP (RFC4341), which formally introduced modular congestion control, 
 \toolname created two grouped nodes to handle the same event: receiving a \texttt{DCCP-Reset} packet. One was created for \texttt{CLOSED}, \texttt{LISTEN} and \texttt{TIMEWAIT} states, for reset code $3$, and a second for the \texttt{REQUEST} and \texttt{RESPOND} states, for reset code $4$. 
These distinctions, missing from the diagram but specified in the RFC text, are made explicit in \toolname's summary (which is elided for brevity).
\section{Related work}\label{sec:related}

For decades, 
RFCs have been
the \emph{de facto} standard for specifying network protocols. 
This reliance on 
prose-based,
informal specifications has motivated a broad range of 
NLP-, neural network-, and LLM-based tools. These efforts have primarily targeted code generation, protocol fuzzing, and attack synthesis. 
More recently, PROSPER~\cite{sharma_prosper_2023} introduced an approach for extracting FSMs from RFCs with the explicit goal of aiding protocol understanding.

\inlinehead{\textbf{Code Generation}} Early efforts such as SAGE~\cite{yen_semi-automated_2021} used semi-automated NLP methods to extract logical forms from RFCs for code generation and ambiguity detection. SAGE notably generated interoperable ICMP implementations, demonstrating feasibility for simple protocols. However, this approach was limited in supporting complex state management (e.g., TCP), relied on user-supplied technical lexicons, and did not provide human-interpretable links to the original RFC text.



\inlinehead{\textbf{Fuzzing and Attack Synthesis}} Shortly after SAGE, a series of contributions 
addressed RFCs from an adversarial perspective, using protocol specifications
to guide fuzzing and synthesize attacks.
\citet{jero_leveraging_2019} introduced an NLP-based approach to extract formal representations of a protocol's packet-space, as well as optional properties to generate test cases for grammar-assisted fuzzing. \citet{pacheco_automated_2022} improved on this with BERT~\cite{bert}-based neural models to extract FSMs, which are then used to seed traces used for attack synthesis. Unlike SAGE, their tool does not require a user-provided technical lexicon; instead, they train embeddings for protocol terminology, as part of model pretraining. 
\citet{zhang_blackbox_2023} similarly leverage BERT to extract intermediate FSMs, but to generate traces for fuzzing. \citet{meng_large_2024} use LLMs directly for the same purpose, relying on the LLM’s training data rather than providing the RFC as input.
\citet{liang_automatic_2024} also use a BERT-based RFC parsing approach; however, they aim to extract \emph{Petri nets} rather than abstract FSMs.

While the majority of these approaches, like ours, extract FSMs from RFCs, their primary goal is to produce intermediate models for automated analysis or testing. Our work differs in three key ways:
First, the FSMs produced by prior work are machine-readable representations intended for trace synthesis, whereas \toolname provides users with human-interpretable summary diagrams for interactive exploration. Second, although our summaries are LLM-extracted, we prioritize \emph{soundness} with respect to the RFC: approximate FSMs may suffice for automation, but for understanding and auditability, correctness is essential. Third, every transition in \toolname is explicitly linked to its supporting RFC text, allowing users to verify and audit the extracted protocol logic.


\inlinehead{\textbf{Knowledge Acquisition}}
Unlike the aforementioned related contributions, \citet{sharma_prosper_2023} describe PROSPER, an LLM-based tool to extract FSMs from RFCs with the explicit goal of human interpretation and knowledge acquisition. PROSPER operates by first selecting and cleaning RFCs, removing metadata and appendices,  and chunking them into 500-line portions. In parallel,  their custom Artifact Miner tool extracts non-textual artifacts such as built-in FSM ASCII diagrams. Each chunk and artifact is then provided to the LLM, which is prompted to output Python code for any protocol-related transitions inferred from the input. Finally, the LLM assembles these into an aggregated FSM model.


In contrast, our approach differs from PROSPER's in three main ways: 
First, every data point in \toolname's summary diagrams is explicitly linked to its RFC source, allowing users to verify soundness and directly audit the origin of each extracted transition.
Second, our pipeline combines RFC partitioning with context-aware LLM queries, ensuring that cross-references and semantically related details from across the RFC are considered together during summary extraction.
Third, whereas PROSPER’s FSMs largely mirror the structure and detail of RFC diagrams, \toolname's summary diagrams reliably recover nearly all diagrammed nodes and transitions (see \Cref{tab:rfc_methods_two}), and consistently capture additional protocol elements documented only in the RFC text. This broader, more semantic summary representation, including conditions, actions, and context, enables richer protocol analysis and is especially valuable for RFCs with incomplete diagrams, such as QUIC. 

\section{Discussion}\label{sec:discussion}


Summary Visualization offers a promising new way to assist both implementers and RFC authors in protocol logic understanding and evolution. By grounding every extracted element in RFC source text, \toolname delivers a level of transparency and auditability not previously possible for protocol visualizations. We hope these results encourage further efforts to bridge the long-standing gap between protocol specifications and real-world implementations. 
Looking ahead, we plan to extend \toolname to integrate data from multiple RFCs, allowing users to compare updates and relationships across protocols. This should help surface ambiguities and contradictions earlier in the RFC life cycle.


\subsection*{Ethical Concerns}
This work raises no ethical concerns. All data consisted of RFC documents and researcher-submitted queries to the OpenAI API, and no private or personally identifiable information was involved at any stage.

\bibliographystyle{ACM-Reference-Format} 
\bibliography{references}

\end{document}